\documentclass[conference]{IEEEtran}
\IEEEoverridecommandlockouts
\usepackage{cite}
\usepackage{amsmath,amssymb,amsfonts}

\usepackage{dblfloatfix}
\usepackage{float}  
\usepackage{algorithmic}
\usepackage{graphicx}
\usepackage{textcomp}
\usepackage{xcolor}
\usepackage{xspace}
\usepackage{soul}
\usepackage{array}
\usepackage{booktabs}
\usepackage{listings}
\usepackage{url}
\usepackage{hyperref}
\usepackage{enumitem}
\usepackage{tikz}
\usetikzlibrary{arrows.meta,positioning}

\makeatletter
\newcommand{\linebreakand}{%
  \end{@IEEEauthorhalign}
  \hfill\mbox{}\par
  \mbox{}\hfill\begin{@IEEEauthorhalign}
}
\makeatother

\def\BibTeX{{\rm B\kern-.05em{\sc i\kern-.025em b}\kern-.08em
    T\kern-.1667em\lower.7ex\hbox{E}\kern-.125emX}}
\begin{document}

\title{RADAgent: Automating RADseq Pipelines with Private Open-Source LLMs}

\author{\IEEEauthorblockN{Qiming Guo}
\IEEEauthorblockA{
\textit{Texas A\&M University-Corpus Christi}\\
Corpus Christi, TX \\
qguo2@islander.tamucc.edu}
\and
\IEEEauthorblockN{Malaka Sudhakara Bava Sukruth Reddy}
\IEEEauthorblockA{
\textit{Texas A\&M University-Corpus Christi}\\
Corpus Christi, TX \\
malakasudhakarab.reddy@tamucc.edu}
\and
\IEEEauthorblockN{Honggang Zhao}
\IEEEauthorblockA{
\textit{Center for Aquaculture Technologies}\\
San Diego, CA \\
hzhao@aquatechcenter.com}
\linebreakand
\IEEEauthorblockN{Ye Wang}
\IEEEauthorblockA{
\textit{Biogen}\\
Cambridge, MA \\
ye.wang@biogen.com}
\and
\IEEEauthorblockN{Chen Pan}
\IEEEauthorblockA{
\textit{The University of Texas at San Antonio}\\
San Antonio, TX \\
chen.pan@utsa.edu}
\and
\IEEEauthorblockN{Wenlu Wang}
\IEEEauthorblockA{
\textit{Texas A\&M University-Corpus Christi}\\
Corpus Christi, TX \\
wenlu.wang@tamucc.edu}
}

\maketitle

\begin{abstract}
Restriction site-associated DNA sequencing (RADseq) is a family of reduced-representation sequencing methods. It uses restriction enzymes to sample subsets of the genome, therefore allowing multiple individuals to be sequenced and genotyped together at low cost. RADseq is widely used for population genomics, evolutionary biology, marker discovery, and the development of SNP panels, but complete analysis still requires expert coordination across multiple tools and processing stages. Traditional RADseq analysis requires expert coordination of heterogeneous bioinformatics tools, intermediate artifacts, quality-control checks, and parameter-dependent decisions across multiple stages. Although individual steps can be scripted, executing the complete workflow still requires human oversight for state tracking, result validation, and failure handling. To automate this expert-driven coordination, we present \textbf{RADAgent}, which introduces a modular agentic method for reliable RADseq workflow execution with an open-weight LLM. RADAgent maps the natural boundaries of the RADseq workflow into 16 independently executable and verifiable agent modules and uses a model-independent runtime for bounded tool execution, on-disk verification, and failure recovery. We evaluate RADAgent on a real population-genomics dataset containing 476 samples from 19 populations. The results demonstrate that RADAgent reliably automates the complete RADseq workflow with an open-weight LLM while keeping genomic data and model execution within the local computing environment. The code is publicly available at \url{https://github.com/wenlu-lab/RADAgent}.

\end{abstract}


\section{Introduction}
Restriction site-associated DNA sequencing (RADseq) recovers thousands of genome-wide markers without requiring prior genomic resources for the study species~\cite{baird2008rad,andrews2016radseq}. It is now a standard approach for SNP discovery and genotyping in non-model organisms, supporting
studies of population structure, introgression, selection, and the development of SNP panels~\cite{andrews2016radseq}. However, converting
raw reads into a reliable marker panel requires chaining multiple steps, each with its own scripts, file formats, and quality control criteria. These parameter choices are consequential, and filtering and
locus-assembly settings alone can substantially change the population genetic inference drawn from the same
reads~\cite{andrews2016radseq,shafer2017bioinformatic}.

RADseq analysis requires coordinating multiple tools and scripts from raw FASTQ reads through BAM processing, variant calling, VCF filtering, and quality control. While individual steps can be scripted, complete automation still requires intelligent validation of intermediate outputs and handling of execution failures. LLM agents provide this capability through tool use and reasoning, but existing hosted platforms may expose genomic data and intermediate results to external services, raising privacy and data-governance concerns.

To address these challenges, we introduce \textbf{RADAgent}, an open-source agentic framework that automates RADseq workflows with locally deployed open-weight LLMs. RADAgent decomposes the workflow into 16 naturally bounded steps, each executed by an isolated sub-agent with step-specific tool access and explicit verification criteria. A key design principle is to separate agent reasoning from workflow control: the LLM interprets step-specific procedures and interacts with bioinformatics tools, while a model-independent runtime verifies the resulting artifacts and governs workflow progression. This modular design bounds agent execution and failure recovery, reducing context contamination, constraining each agent's action space, and preventing errors from propagating across pipeline stages. Together, these mechanisms enable smaller open-weight LLMs to reliably execute long-running scientific workflows without relying on model self-reports to determine execution correctness.

\begin{figure*}[!t]
    \centering
   \includegraphics[width=1\linewidth]{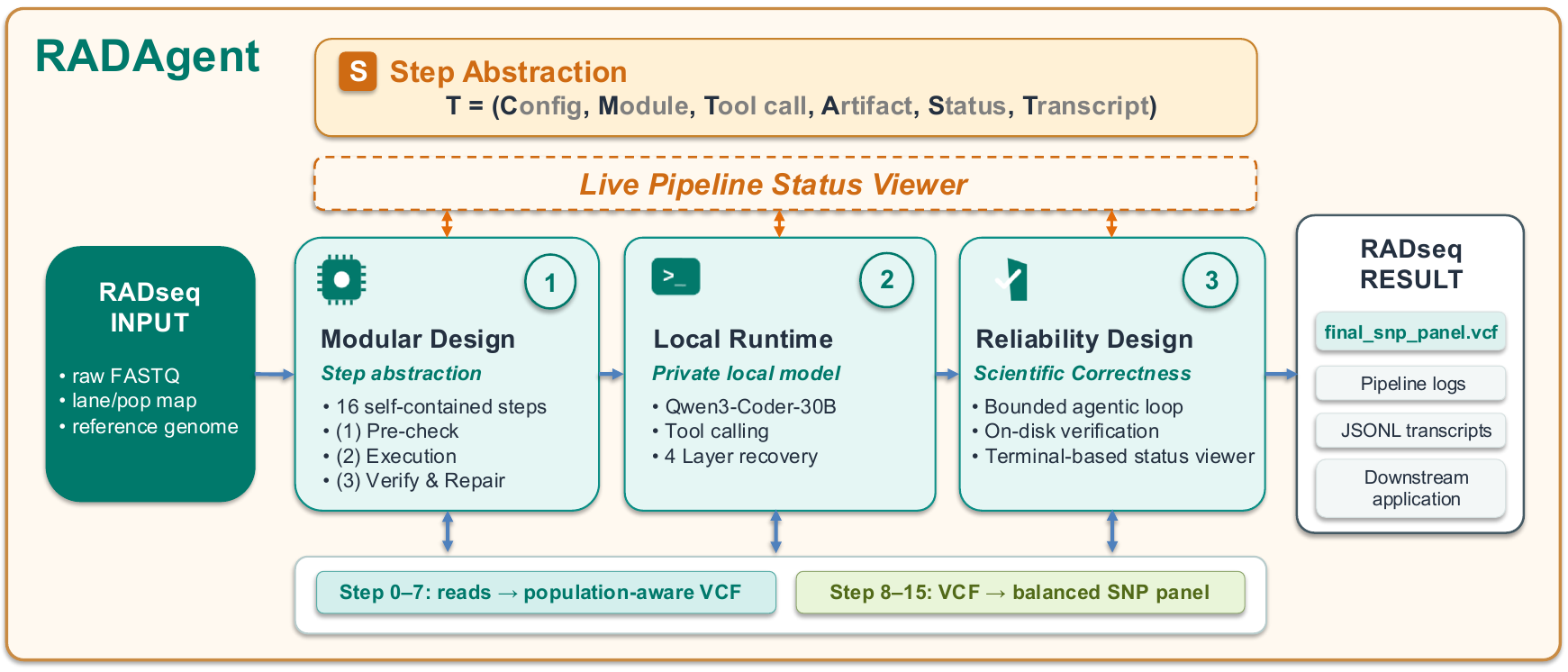}
    \caption{RADAgent Overview}
    \label{fig:overview}
\end{figure*}

RADAgent was evaluated on a real population-genomics dataset containing 476 crayfish RADseq samples from 19 populations ~\cite{adams2024genomic}. The results demonstrate that the proposed agentic design can reliably automate the complete RADseq workflow with a 30B open-weight LLM, while keeping genomic data and model execution within the local computing environment. Our analysis further examines how runtime verification and recovery contribute to reliable execution and identifies the remaining failure modes of smaller open-weight models.
More broadly, RADAgent aims to reduce the technical burden of complex bioinformatics workflows, allowing researchers to focus on higher-value scientific tasks.

\section{Related Work}

RADseq denotes a family of protocols that differ in enzyme choice, size selection, and library preparation. These include original
RAD ~\cite{miller2007rad,baird2008rad}, genotyping-by-sequencing (GBS) ~\cite{elshire2011gbs}, double-digest RAD (ddRAD)~\cite{peterson2012ddrad}, 2b-RAD ~\cite{wang2012twobrad}, ezRAD ~\cite{toonen2013ezrad}, and RAD capture ~\cite{ali2016rapture}.
Following Andrews et al. ~\cite{andrews2016radseq}, we use RADseq for the family as a whole, because all members use restriction enzymes to define which loci are sequenced. Analysis of these data is supported by
established pipelines such as Stacks ~\cite{catchen2013stacks,
rochette2019stacks2} and dDocent ~\cite{puritz2014ddocent}, which assemble
reads into loci de novo or against a reference genome and call variants across individuals. These pipelines
automate individual stages, but the analyst must still chain them and choose filtering parameters by hand. Such errors are difficult to detect, because a finished run can yield a marker set that is subtly wrong and intermediate artifacts are rarely inspected once a step reports success. The same stepwise structure underlies whole-genome (WGS) and pooled sequencing(pool-seq) data ~\cite{schlotterer2014poolseq}, so the modular design presented here may extend beyond RADseq.

LLM agents are increasingly used to automate complex scientific tasks by combining language-model reasoning with external tools and multi-stage workflows~\cite{schmidgall2025agentlab,lu2024aiscientist,bran2024chemcrow,ding2025scitoolagent}. Related work has also advanced agents that act inside computing environments through interleaved reasoning and tool use~\cite{yao2023react,wang2024codeact}, including repository-level software engineering~\cite{jimenez2024swebench,yang2024sweagent,wang2025openhands} and constrained localize--repair--validate pipelines~\cite{zhang2024autocoderover,xia2025agentless}.

Within bioinformatics and biomedical research, agents increasingly support workflow planning, tool selection, and multi-stage analysis~\cite{zhou2024autoba,mehandru2025bioagents,huang2025biomni}. However, reliable execution of long, artifact-producing workflows remains challenging because model-generated status reports alone cannot establish the validity of intermediate scientific outputs. This distinction is critical for RADseq, where each step produces artifacts whose validity must be established independently of the model's self-report. RADAgent therefore adopts the constrained-pipeline philosophy and moves verification, loop control, and recovery into the runtime.

\section{Method}
We introduce \textbf{RADAgent}, which provides a new agentic method for executing expert-defined scientific workflows with locally deployed open-weight LLMs. RADAgent addresses the reliability limitations of using smaller self-hosted models for long-horizon workflow execution by transforming a RADseq procedure into modular agent tasks with explicit execution boundaries, while moving tool control, verification, and failure recovery into a model-independent runtime. The following sections describe the modular agentic design, local execution runtime, and reliability mechanisms that realize this method.

\begin{table*}[t]
\centering
\caption{The sixteen steps that constitute the RADseq pipeline. Each step wraps a single tool boundary and produces a single named artifact.}
\label{tab:pipeline}
\renewcommand{\arraystretch}{1.15}
\begin{tabular}{@{}r l l l l@{}}
\toprule
\textbf{N} & \textbf{Step} & \textbf{Tool / Library} & \textbf{Key Parameters} & \textbf{Primary Output} \\
\midrule
0  & rad-0-lane-info     & Bash, perl                  &                                    & \texttt{01-info\_files/lane\_info.txt} \\
1  & rad-1-cutadapt      & cutadapt + GNU parallel     & $e=0.2$, $m=50$ bp                   & \texttt{02-raw/trimmed/*.fastq.gz} \\
2  & rad-2-radtags       & Stacks process\_radtags     & enzyme \texttt{sbfI}, $-c~-r$, trim 120 bp & \texttt{03-samples/\{lane\}/*} \\
3  & rad-3-rename        & Bash, awk, perl, ln/cat     & merge or link multi-lane samples     & \texttt{04-all\_samples/*.fq.gz} \\
4  & rad-4-popmap        & Bash, awk                   & population $\rightarrow$ ID map      & \texttt{01-info\_files/population\_map.txt} \\
5  & rad-5-bwa           & BWA mem + samtools          & $-q~10$ filter, sorted+indexed       & \texttt{04-all\_samples/*.sorted.bam} \\
6  & rad-6-gstacks       & Stacks gstacks              & \texttt{--max-clipped 0.1}           & \texttt{05-stacks/catalog.*} \\
7  & rad-7-populations   & Stacks populations          & $p=1$, $r=0.6$, ordered VCF          & \texttt{05-stacks/populations.snps.vcf} \\
8  & rad-8-vcf-filter    & Python, vcftools, plink     & $m=4$, $p=80$, $x=0$, MAS=1, 94 chr  & \texttt{filtered\_m4\_p80\_x0\_S1\_chr.recode.vcf} \\
9  & rad-9-snp-dup-hwe   & Python, R, plink            & $F_{is}{<}{-}0.4$; HWE $p{<}10^{-3}$ & \texttt{*.singleton.hwe.recode.vcf} \\
10 & rad-10-ld-clump     & R bigsnpr, plink            & $r^2{<}0.2$, 10-SNP window           & \texttt{*.LD.recode.vcf} \\
11 & rad-11-remove-atgc  & Python                      & strand-ambiguous SNPs removal           & \texttt{*.LD.no\_AT\_CG.vcf} \\
12 & rad-12-maf-filter   & vcftools                    & MAF $\geq 0.1$                       & \texttt{*.maf10.LD.no\_AT\_CG.recode.vcf} \\
13 & rad-13-flanking     & Python                      & $\leq 3$ SNPs and $\leq 1$ indel within $\pm 150$ bp & \texttt{filtered\_flanking.vcf} \\
14 & rad-14-blast-map    & makeblastdb, blastn, perl   & $e{<}10^{-10}$, single best hit      & \texttt{blast\_filtered\_snps.vcf} \\
15 & rad-15-even-dist    & Python                      & target $\sim$1 SNP / 10 Mb           & \texttt{final\_snp\_panel.vcf} \\
\bottomrule
\end{tabular}
\end{table*}

\subsection{Modular Agentic Design}

A RADseq workflow naturally decomposes into a sequence of tool-bounded steps. Each step consumes a well-defined input artifact, runs a specific bioinformatics tool or filtering script, and produces an intermediate output that becomes the input to the next step. 
Instead of asking a self-hosted open-weight model to manage the entire workflow in a single long context, RADAgent assigns each pipeline step to an independent module with explicit inputs, outputs, scripts, verification criteria, and recovery rules. This modular structure reduces context contamination from long bioinformatics logs, constrains each sub-agent's action space through a step-specific tool allowlist, and localizes failures to a single pipeline
boundary. \textbf{For scientific workflows, this is critical because each intermediate artifact must be traceable and verifiable before it is passed to downstream analysis.} RADAgent operationalizes these natural workflow boundaries as agent execution boundaries, allowing each scientific operation to be independently executed, verified, and recovered before its output is propagated downstream.

As shown in Figure~\ref{fig:overview} and Table~\ref{tab:pipeline}, RADAgent decomposes the pipeline into 16 sequential steps: Steps 0--7 transform raw reads into a population-aware VCF, and Steps 8--15 refine it into a balanced SNP panel. Each step follows a pre-check--execution--verification cycle, and the orchestrator advances only after its explicit verification criteria are satisfied.

\textbf{Step Abstraction:}
RADAgent represents each workflow step as a self-contained execution object,
\[
T=(\text{Config},\text{Module},\text{Tool Call},\text{Artifact},\text{Status},\text{Transcript}),
\]
where \textit{Config} defines the step-specific procedure and constraints, \textit{Module} provides the isolated agent context, and \textit{Tool Call}, \textit{Artifact}, \textit{Status}, and \textit{Transcript} capture its execution and observable state. This abstraction separates agent reasoning from workflow state, allowing each step to be independently executed, verified, and traced before the orchestrator advances.

Each step is executed through a common runtime protocol consisting of pre-check, execution, and verify--repair phases. \textbf{Pre-check} verifies that inputs exist and tool dependencies resolve on \texttt{PATH}, removes prior partial outputs, and computes all expected counts \emph{from the live filesystem} rather than as constants:
$
N_{\text{exp}} \;=\; \bigl\lvert \{\, f : f \in \mathtt{02\text{-}raw/*\_1.fastq.gz} \,\} \bigr\rvert
$. 
\textbf{Execute} invokes the wrapped script and records output to a timestamped log. 
\textbf{Verify \& Repair} validates expected artifacts, output counts, non-empty files, tool success markers, and step-specific QC thresholds, including a cutadapt per-sample pass rate $\geq 90\%$ and BWA mapping rate $\geq 95\%$. The model's self-reported status is not trusted; the model-independent on-disk verifier checks the resulting artifacts against the live filesystem, confirming valid outputs or routing failures through the recovery layers.

\subsection{Local Runtime}

\subsubsection{Runtime Design}

The local runtime implements the execution semantics of the Step Objects while keeping tool execution and workflow control outside the LLM. RADAgent runtime is implemented as a single long-lived local process. Because the 30B open-weight model occupies approximately 60~GB of GPU memory in bf16, reloading the model for each pipeline step would be impractical.
Instead, the runtime loads the model once and reuses the same resident instance across the orchestrator, step-level sub-agents, and the repair agent. Each agent creates an isolated context rather than a new operating-system process with its own message history, transcript, and step-specific tool allowlist, while sharing the same loaded model. Thus, switching between the orchestrator and an agent only changes the message context fed to the model, not the model weights in GPU memory.

\subsubsection{Tool Calling}

Because the self-hosted model produces plain text rather than platform-managed (e.g., Claude) structured tool calls, RADAgent implements a lightweight tool-calling layer that injects tool schemas into the prompt and parses the model's structured text into executable calls. Generation stops after each complete call; the runtime executes the requested tool and returns its actual output in the next turn, preventing fabricated tool results while keeping execution outside model control.

\subsubsection{Verify \& Repair Loop}
When verification fails, RADAgent expands the scope of intervention through a bounded recovery hierarchy rather than allowing unrestricted autonomous debugging. We layer four independent recovery mechanisms: \textbf{Layer~1} is the in-step self-heal; \textbf{Layer~2} is an in-step retry of the execute phase immediately after a Layer~1 fix; \textbf{Layer~3} is an orchestrator-level retry that re-invokes the entire step once from a clean state when it returns failure status; and \textbf{Layer~4} is escalation to the repair agent. The orchestrator advances to the next step only when the current step returns a success status, and logs the outcome of every layer.

\begin{table}[!h]
\centering
\caption{Fix-level progression used by the autonomous debugger. Levels are tried lowest-first; exactly one is applied per invocation.}
\label{tab:fixmatrix}
\renewcommand{\arraystretch}{1.15}
\begin{tabular}{@{}r l l@{}}
\toprule
\textbf{Lvl} & \textbf{Action class} & \textbf{Typical categories} \\
\midrule
1 & chmod, mkdir, rm stale          & ENV \\
2 & edit env vars, paths, parameters & ENV, TOOL \\
3 & repair script (syntax, hardcode) & SCR \\
4 & regenerate runtime script        & SCR \\
5 & remove corrupt outputs, re-run   & DAT, DEP \\
6 & swap tool path (e.g., parallel $\rightarrow$ for-loop) & TOOL \\
\bottomrule
\end{tabular}
\end{table}

\textbf{Bounded Repair Agent for Scientific Workflow} The final recovery layer is an autonomous repair agent. 
Unlike step modules, the repair agent is granted write access so that it can repair recoverable runtime failures. When invoked with a step number $N$ and an error excerpt, it (i) globs and reads the failed step's prompt, extracting its expected inputs, outputs, scripts, tools, and verification rules; (ii) reconstructs system state from the pipeline log, the step's own log, disk/memory/CPU, and any zombie processes; (iii) classifies the fault into exactly one of six categories with a stated confidence, \textsc{ENV} (missing binary, permissions), \textsc{RES} (disk or memory exhaustion), \textsc{DEP} (upstream file missing or wrong count), \textsc{DAT} (corrupt or truncated VCF/BAM), \textsc{SCR}
(script syntax or hardcoded path), \textsc{TOOL} (R/Python/segfault/version); (iv)
applies exactly one fix from the six-level progression in Table~\ref{tab:fixmatrix}; and (v) re-runs the step's own verification rules and reports \texttt{FIX\_APPLIED} or \texttt{UNFIXABLE}. 
By design, the repair agent is restricted from modifying protected inputs, including raw FASTQ files, the reference genome, project metadata, files outside the project root, or system-level software.
The single-repair rule forces the agent to commit to one hypothesis at a time, makes each repair easy to audit, and limits the blast radius of automated edits. Such constraints are particularly important for smaller open-weight models, which may be less reliable at termination control and fault diagnosis in long-horizon workflows.

\subsection{Reliability Design for Scientific Pipelines}

\subsubsection{Bounded Agentic Loop}

RADAgent executes each step through a bounded model--tool interaction loop rather than an open-ended agent process. At each turn, the runtime renders the current message history, obtains one model response, executes any requested tool calls, and feeds the real tool outputs back into the next turn. When no tool call is returned, the runtime parses the step's structured status report and returns control to the orchestrator. This design keeps tool execution outside the model while allowing the model to iteratively inspect files, run commands, and verify step completion. We design several robustness mechanisms to make the loop reliable for a self-hosted open-weight model.

First, the runtime bounds the maximum number of tool calls per step (200 by default), preventing unbounded execution when the model fails to terminate. Second, because long bioinformatics logs can quickly exhaust the context window (32{,}768 tokens by default), older tool outputs are compacted while preserving the system prompt, the original task request, and the most recent interaction history. Third, the runtime detects repeated identical tool calls and prompts the model to stop tool use and emit the required structured summary. Finally, each step must end with a machine-readable status marker, which the orchestrator uses to decide whether to advance, retry, or escalate to the repair agent. Together, these adaptations make the open-model loop bounded, auditable, and recoverable.

\subsubsection{Model-Independent On-Disk Verification}\label{sec:ondisk}

A key reliability principle in RADAgent is to separate agent execution from workflow-state validation. The model's textual status report is therefore not treated as the sole indicator of step completion. Because an open-weight model may report failure after producing valid outputs or report success before all required checks have passed, RADAgent uses model-independent on-disk verification to determine whether the step's completion criteria are satisfied.

After each sub-agent finishes, the runtime performs an on-disk check for the corresponding step. These checks are step-specific and are based on the actual files generated by the pipeline. For example, the verifier checks whether expected FASTQ, BAM, VCF, or final-panel files exist, are non-empty, and satisfy basic structural requirements such as the presence of variant records in a VCF. If the model reports failure but the expected outputs are present and valid on disk, the runtime overrides the step status to \texttt{SUCCESS} and logs the override. Otherwise, the failure is routed to the recovery layers.

For auditability, RADAgent records model messages, tool calls, outputs, and repair events as local append-only JSONL transcripts that can be inspected through the live terminal viewer, \texttt{./rad view}. Because both execution and transcripts remain local, runs can be audited without relying on external services.

\begin{table}[!h]
\centering
\caption{Per-population sample counts entering the HWE filter.}
\label{tab:popqc}
\renewcommand{\arraystretch}{1.05}
\begin{tabular}{@{}l r l r@{}}
\toprule
\textbf{Population} & \textbf{N} &
\textbf{Population} & \textbf{N} \\
\midrule
Apartment1 & 21 & Hotel1   & 77 \\
Apartment2 &  5 & Hotel2   & 10 \\
Apartment3 &  5 & Hotel3   & 18 \\
Drain1     & 26 & Hotel4   & 17 \\
EastGC10   & 24 & Private1 & 20 \\
EastGC11   & 23 & WestGC1  & 49 \\
EastGC13   & 30 & WestGC2  & 24 \\
EastGC2    & 47 & WestGC3  &  2 \\
EastGC4    & 33 & WestGC6  & 23 \\
EastGC7    & 22 & \textit{Total} & \textbf{476} \\
\bottomrule
\end{tabular}
\end{table}

\section{Experiments}

\subsection{Dataset and Environment}\label{sec:dataset}
We use a real population-genomics study of crayfish~\cite{adams2024genomic} (\emph{Procambarus clarkii}):
476 paired-end Illumina samples (952 FASTQ files, $\sim$101~GB) drawn from 19
populations, against a reference of 94 NCBI scaffolds
(\texttt{NC\_091150.1}--\texttt{NC\_091243.1}) used for both BWA alignment and
downstream BLAST validation. The agent runs on a single Linux server with one
NVIDIA A100 80~GB PCIe GPU (81{,}920~MiB, compute capability 8.0). 
The per-population counts are shown in Table~\ref{tab:popqc}.

\subsection{On-Disk Verification}
The on-disk verifier is the runtime's substitute for the open model's unreliable
self-report, and it fires often. Mining 16 orchestrator runs, it overrode a
non-\texttt{SUCCESS} self-report whose outputs were nonetheless valid on disk
\textbf{15 times} (Table~\ref{tab:override}): $\sim$0.94 overrides per run, 8.9\% of
all 169 step executions, with at least one override in 11 of 16 runs (69\%). Nine
overrides were the model returning \texttt{FAILURE}/\texttt{UNFIXABLE} for valid
work; six were the model exhausting its tool-call budget on work that had
\emph{already completed} on disk. Conservatively counting one prevented Layer-3
re-run per override, the verifier saved an estimated \textbf{4.58~h} of GPU compute
across those runs. The ablation gate confirms the mechanism: with
\texttt{RAD\_DISABLE\_CROSSCHECK=1} each of the 15 events would instead have
surfaced as a step failure (needless retry or halt).

\begin{table}[H]
\centering
\caption{On-disk verification overrides across 16 orchestrator runs. ``Fail'' denotes model-reported failure despite valid outputs; ``Budget'' denotes completed work that reached the 200-call limit.}
\label{tab:override}
\renewcommand{\arraystretch}{1.1}
\begin{tabular}{@{}r l r r r@{}}
\toprule
\textbf{N} & \textbf{Step} & \textbf{Overrides} & \textbf{Fail} & \textbf{Budget} \\
\midrule
1 & rad-1-cutadapt    & 3 & 2 & 1 \\
2 & rad-2-radtags     & 4 & 3 & 1 \\
3 & rad-3-rename      & 2 & 0 & 2 \\
6 & rad-6-gstacks     & 2 & 1 & 1 \\
7 & rad-7-populations & 3 & 3 & 0 \\
8 & rad-8-vcf-filter  & 1 & 0 & 1 \\
\midrule
\textbf{Total} & & \textbf{15} & \textbf{9} & \textbf{6} \\
\bottomrule
\end{tabular}
\end{table}

The verifier is deliberately a \emph{structural/liveness} gate rather than a deep
correctness check, and we characterize that boundary honestly. Injecting five
corruption types into the tail VCFs yields a confusion matrix of
TP\,$=$\,9, FP\,$=$\,12, FN\,$=$\,2, TN\,$=$\,12 (precision 0.43): it reliably
catches zero-byte, header-only, and missing/wrong-count outputs but passes some
mid-line truncations and wrong-format files. A broader synthetic battery (six
fault types $\times$ 16 steps) put its recall at 39/87 (44.8\%) on applicable
faults. These limitations are by design: deep numerical correctness is instead checked offline against committed per-step reference fixtures, which caught all 12 structurally valid but numerically wrong files.

\begin{table}[H]
\centering
\caption{Effectiveness of loop-detection intervention across workflow steps.}
\label{tab:nudge}
\renewcommand{\arraystretch}{1.1}
\begin{tabular}{@{}l r r r r r@{}}
\toprule
\textbf{Step} & \textbf{Detected} & \textbf{Resolved} & \textbf{Cap} & \textbf{Succ.} & \textbf{Fail} \\
\midrule
rad-1-cutadapt    & 2 & 1 & 1 & 1 & 0 \\
rad-3-rename      & 3 & 3 & 0 & 3 & 0 \\
rad-4-popmap      & 1 & 1 & 0 & 1 & 0 \\
rad-5-bwa         & 5 & 4 & 1 & 3 & 1 \\
rad-6-gstacks     & 3 & 3 & 0 & 0 & 3 \\
\midrule
\textbf{Total}    & \textbf{14} & \textbf{12} & \textbf{2} & \textbf{8} & \textbf{4} \\
\bottomrule
\end{tabular}
\end{table}

\subsection{Loop Detection}

The open model lacks an innate ``task-complete'' instinct and instead repeats a
tool call. The nudge detects this and forces a summary. Across the transcript
corpus the nudge fired \textbf{14 times} in 6 sessions and stopped the loop in
\textbf{12 of 14 cases} (85.7\%); the two that still hit the call cap were
absorbed downstream by the budget-exhaustion path of the on-disk verifier
(Table~\ref{tab:nudge}). A controlled ablation isolates the effect: replaying the
real runtime loop against a synthetic ``responsive'' client that loops until
nudged, $k{=}0$ nudges runs to the cap, whereas $k{\geq}1$ terminates cleanly at
turn~4; a synthetic ``resistant'' client loops to the cap regardless of $k$,
confirming the nudge helps recoverable loops without masking genuinely stuck ones.
Cheap deterministic steps never repeated a call and fired zero nudges.

\begin{table}[H]
\centering
\caption{Context-trimming events observed across 267 execution transcripts.}
\label{tab:trim}
\renewcommand{\arraystretch}{1.1}
\begin{tabular}{@{}l r r r@{}}
\toprule
\textbf{Step} & \textbf{Events} & \textbf{Tokens Dropped} & \textbf{Peak Input} \\
\midrule
rad-6-gstacks       & 22 & 193{,}177 & 115{,}495 \\
Other steps         & 2  & 1{,}001   & $<$40{,}000 \\
\midrule
\textbf{Total}      & \textbf{24} & \textbf{194{,}178} & \\
\bottomrule
\end{tabular}
\end{table}

\begin{table*}[!t]
\centering
\caption{Representative faults and the recovery mechanism that resolved each. The last two rows are failure modes specific to driving an open model.}
\label{tab:failures}
\renewcommand{\arraystretch}{1.15}
\begin{tabular}{@{}l l l l@{}}
\toprule
\textbf{Step} & \textbf{Symptom} & \textbf{Diagnosis} & \textbf{Resolution} \\
\midrule
1 & \texttt{parallel: command not found}      & GNU parallel missing on \texttt{PATH}      & L1: sequential bash fallback \\
2 & \texttt{Permission denied} on script      & Lost \texttt{+x} bit after git checkout    & L1: \texttt{chmod +x} self-heal \\
3 & Right-mate FASTQ pairs missing            & Prior \texttt{kill -9} during rename       & L4: debugger regenerates \texttt{*.2.fq.gz} links \\
5 & BWA wall time exceeds expectation         & Background launched without \texttt{nohup} & L1: relaunch under \texttt{nohup}, adaptive poll \\
6 & gstacks output VCF zero bytes             & BAM index missing for one sample           & L4: debugger reindexes via \texttt{samtools index} \\
9 & R \texttt{plink} \texttt{--hwe} CRLF error & Script saved on macOS pulled to Linux     & L4: debugger normalizes line endings \\
10 & bigsnpr \texttt{snp\_readBed} stale .bk   & Backing file from previous run             & L1: remove stale \texttt{.bk}, retry \\
any & \texttt{FAILURE} reported, outputs valid & Open model under-reports success           & Runtime on-disk cross-check overrides to SUCCESS \\
any & Identical tool call repeats indefinitely & Open model lacks task-complete instinct    & Runtime loop-detection nudge forces summary \\
\bottomrule
\end{tabular}
\end{table*}

\begin{table*}[!t]
\centering
\caption{Fault-injection evaluation of the four-layer recovery mechanism (on-disk verifier disabled to isolate recovery-layer behavior).}
\label{tab:routing}
\renewcommand{\arraystretch}{1.15}
\begin{tabular}{@{}l c l l c c@{}}
\toprule
\textbf{Injected Fault} & \textbf{Step} & \textbf{Expected} & \textbf{Actual Recovery} & \textbf{Final Status} & \textbf{Output OK} \\
\midrule
Missing dependency, backup available & 9 & L4 & L4 debugger & SUCCESS & yes \\
Missing dependency, no backup        & 9 & L4 & Retry loop; no escalation & TIMEOUT & n/a \\
Script syntax error                   & 8 & L4 & L1 self-repair & SUCCESS & yes \\
Stale artifact                        & 9 & L1 & L1 removal & SUCCESS & yes \\
Corrupt VCF                           & 9 & L4 & L1 wrong-file substitution & SUCCESS & \textbf{no} \\
\bottomrule
\end{tabular}
\end{table*}

\subsection{Context Management}

Because in-process \texttt{transformers} re-encodes the full prompt every turn,
the trimmer is not a safety net but a routinely exercised
mechanism. A subtlety: grepping transcripts for the trim placeholder always returns
zero, because the placeholder is mutated in memory \emph{after} the transcript line
is written; trimming must instead be detected from a drop in recorded
\texttt{input\_tokens} between consecutive turns (untrimmed conversations only
append, so the count is monotonic). Scanning 267 production transcripts, 7
transcripts across 5 sessions trimmed, totaling \textbf{24 trim events} and
\textbf{194{,}178 tokens dropped} (Table~\ref{tab:trim}); the heaviest single
sub-agent (a \texttt{rad-6-gstacks} run) peaked at 115{,}495 input tokens and
trimmed 11 times, dropping 101{,}391 tokens, well past the default window. A budget sweep over 208 sessions found a peak untrimmed conversation
of 149{,}750 tokens and that 7 sessions would trim at the default; live runs at
8{,}192, 32{,}768, and an aggressive 5{,}120-token budget all completed with valid
step-8 outputs, confirming the trimmed conversation still carries enough state to
finish the task.

\begin{table}[H]
\centering
\caption{Retrospective audit of nine debugger invocations and their recovery outcomes.}
\label{tab:debugger}
\renewcommand{\arraystretch}{1.05}
\begin{tabular}{@{}c l l c c@{}}
\toprule
\textbf{Step} & \textbf{Fault Type} & \textbf{Result} & \textbf{Fix Lvl.} & \textbf{Valid.} \\
\midrule
5 & Dependency & UNFIXABLE     & --- & --- \\
6 & Script     & FIX\_APPLIED  & 3   & pass \\
5 & Resource   & FIX\_APPLIED  & 1   & --- \\
6 & Dependency & UNFIXABLE     & --- & --- \\
6 & Dependency & UNFIXABLE     & --- & --- \\
5 & Resource   & FIX\_APPLIED  & 1   & pass \\
6 & Data       & FIX\_APPLIED  & 1   & pass \\
6 & Data       & FIX\_APPLIED  & 1   & pass \\
9 & Dependency & FIX\_APPLIED  & 1   & pass \\
\midrule
\bottomrule
\end{tabular}
\end{table}

\subsection{Failure Recovery}

Table~\ref{tab:failures} summarizes representative execution faults and their
recovery mechanisms, including both bioinformatics workflow failures and
open-model-specific failure modes. To evaluate the recovery cascade more
systematically, we ran a five-fault injection battery (Table~\ref{tab:routing})
with the on-disk verifier disabled to isolate the behavior of individual recovery
layers. A stale artifact self-healed at Layer~1 as designed, while a script syntax
error was repaired in-step without escalation. In contrast, a corrupt VCF
triggered an incorrect Layer~1 file substitution, producing a false-success
output. A missing dependency with no available backup also exposed an orchestrator
gap: the step was retried five times without escalating to the Layer~4 debugger.
Overall, only one of the five injected faults reached Layer~4, revealing
limitations in the current escalation mechanism.

A retrospective audit of every historical \texttt{rad-debugger} invocation
(Table~\ref{tab:debugger}) gives a clearer picture of the debugger itself: 9
invocations, of which 6 reported \texttt{FIX\_APPLIED} (66.7\%) and 3
\texttt{UNFIXABLE}, with 5 validations passing. Every invocation (9/9) read the
failing step's execution prompt before acting, and the applied fixes clustered
at the lowest levels (mean fix level 1.4), consistent with the single-fix
discipline (4 of 6 fixes were strictly single-fix-compliant). In 2 of the 9
invocations, both ultimately \texttt{UNFIXABLE} and from early development, the
model attempted a protected-boundary write: rebuilding the BWA index under
\texttt{08-genome/} when the reference was missing, or writing a population map
into a separate project checkout's \texttt{01-info\_files/}. Neither touched the
authoritative dataset; these cases motivate the narrow tool grants, single-fix
rule, and per-run transcripts used to constrain and audit smaller open models.

\section{Conclusion}
We presented \textbf{RADAgent}, an open-source agentic framework for reliable RADseq automation with locally deployed open-weight LLMs. Evaluation on 476 real RADseq samples demonstrated reliable end-to-end execution while keeping genomic data and model execution local. Overall, RADAgent provides a practical framework for agentic workflow automation with local data control, further demonstrating the potential of LLM agents to advance scientific research and discovery.

% \section*{Acknowledgment}
% This work is partially supported by NSF Award No. 2112631 and the TAMUS REF award.

\clearpage
\bibliographystyle{IEEEtran}
\bibliography{ref}

@article{adams2024genomic,
  title={Genomic data characterize reproductive ecology patterns in Michigan invasive red swamp crayfish (Procambarus clarkii)},
  author={Adams, Nicole E and Homola, Jared J and Sard, Nicholas M and Nathan, Lucas R and Roth, Brian M and Robinson, John D and Scribner, Kim T},
  journal={Evolutionary Applications},
  volume={17},
  number={9},
  pages={e70007},
  year={2024},
  publisher={Wiley Online Library}
}

@article{miller2007rad,
  author  = {Miller, Michael R. and Dunham, Joseph P. and Amores, Angel and Cresko, William A. and Johnson, Eric A.},
  title   = {Rapid and cost-effective polymorphism identification and genotyping using restriction site associated {DNA} ({RAD}) markers},
  journal = {Genome Research},
  volume  = {17},
  number  = {2},
  pages   = {240--248},
  year    = {2007},
  doi     = {10.1101/gr.5681207}
}

@article{baird2008rad,
  author  = {Baird, Nathan A. and Etter, Paul D. and Atwood, Tressa S. and Currey, Mark C. and Shiver, Anthony L. and Lewis, Zachary A. and Selker, Eric U. and Cresko, William A. and Johnson, Eric A.},
  title   = {Rapid {SNP} discovery and genetic mapping using sequenced {RAD} markers},
  journal = {PLoS ONE},
  volume  = {3},
  number  = {10},
  pages   = {e3376},
  year    = {2008},
  doi     = {10.1371/journal.pone.0003376}
}

@article{elshire2011gbs,
  author  = {Elshire, Robert J. and Glaubitz, Jeffrey C. and Sun, Qi and Poland, Jesse A. and Kawamoto, Ken and Buckler, Edward S. and Mitchell, Sharon E.},
  title   = {A robust, simple genotyping-by-sequencing ({GBS}) approach for high diversity species},
  journal = {PLoS ONE},
  volume  = {6},
  number  = {5},
  pages   = {e19379},
  year    = {2011},
  doi     = {10.1371/journal.pone.0019379}
}

@article{peterson2012ddrad,
  author  = {Peterson, Brant K. and Weber, Jesse N. and Kay, Emily H. and Fisher, Heidi S. and Hoekstra, Hopi E.},
  title   = {Double digest {RADseq}: an inexpensive method for de novo {SNP} discovery and genotyping in model and non-model species},
  journal = {PLoS ONE},
  volume  = {7},
  number  = {5},
  pages   = {e37135},
  year    = {2012},
  doi     = {10.1371/journal.pone.0037135}
}

@article{wang2012twobrad,
  author  = {Wang, Shi and Meyer, Eli and McKay, John K. and Matz, Mikhail V.},
  title   = {2b-{RAD}: a simple and flexible method for genome-wide genotyping},
  journal = {Nature Methods},
  volume  = {9},
  number  = {8},
  pages   = {808--810},
  year    = {2012},
  doi     = {10.1038/nmeth.2023}
}

@article{toonen2013ezrad,
  author  = {Toonen, Robert J. and Puritz, Jonathan B. and Forsman, Zac H. and Whitney, Jonathan L. and Fernandez-Silva, Iria and Andrews, Kimberly R. and Bird, Christopher E.},
  title   = {{ezRAD}: a simplified method for genomic genotyping in non-model organisms},
  journal = {PeerJ},
  volume  = {1},
  pages   = {e203},
  year    = {2013},
  doi     = {10.7717/peerj.203}
}

@article{ali2016rapture,
  author  = {Ali, Omar A. and O'Rourke, Sean M. and Amish, Stephen J. and Meek, Mariah H. and Luikart, Gordon and Jeffres, Carson and Miller, Michael R.},
  title   = {{RAD} capture ({Rapture}): flexible and efficient sequence-based genotyping},
  journal = {Genetics},
  volume  = {202},
  number  = {2},
  pages   = {389--400},
  year    = {2016},
  doi     = {10.1534/genetics.115.183665}
}

@article{andrews2016radseq,
  author  = {Andrews, Kimberly R. and Good, Jeffrey M. and Miller, Michael R. and Luikart, Gordon and Hohenlohe, Paul A.},
  title   = {Harnessing the power of {RADseq} for ecological and evolutionary genomics},
  journal = {Nature Reviews Genetics},
  volume  = {17},
  number  = {2},
  pages   = {81--92},
  year    = {2016},
  doi     = {10.1038/nrg.2015.28}
}

@article{catchen2013stacks,
  author  = {Catchen, Julian and Hohenlohe, Paul A. and Bassham, Susan and Amores, Angel and Cresko, William A.},
  title   = {Stacks: an analysis tool set for population genomics},
  journal = {Molecular Ecology},
  volume  = {22},
  number  = {11},
  pages   = {3124--3140},
  year    = {2013},
  doi     = {10.1111/mec.12354}
}

@article{puritz2014ddocent,
  author  = {Puritz, Jonathan B. and Hollenbeck, Christopher M. and Gold, John R.},
  title   = {{dDocent}: a {RADseq}, variant-calling pipeline designed for population genomics of non-model organisms},
  journal = {PeerJ},
  volume  = {2},
  pages   = {e431},
  year    = {2014},
  doi     = {10.7717/peerj.431}
}

@article{shafer2017bioinformatic,
  author  = {Shafer, Aaron B. A. and Peart, Claire E. and Tusso, Sergio and Maayan, Inbar and Brelsford, Alan and Wheat, Christopher W. and Wolf, Jochen B. W.},
  title   = {Bioinformatic processing of {RAD}-seq data dramatically impacts downstream population genetic inference},
  journal = {Methods in Ecology and Evolution},
  volume  = {8},
  number  = {8},
  pages   = {907--917},
  year    = {2017},
  doi     = {10.1111/2041-210X.12700}
}

@article{schlotterer2014poolseq,
  author  = {Schl{\"o}tterer, Christian and Tobler, Raymond and Kofler, Robert and Nolte, Viola},
  title   = {Sequencing pools of individuals --- mining genome-wide polymorphism data without big funding},
  journal = {Nature Reviews Genetics},
  volume  = {15},
  number  = {11},
  pages   = {749--763},
  year    = {2014},
  doi     = {10.1038/nrg3803}
}

@article{rochette2019stacks2,
  author  = {Rochette, Nicolas C. and Rivera-Col{\'o}n, Angel G. and Catchen, Julian M.},
  title   = {Stacks 2: Analytical methods for paired-end sequencing improve {RADseq}-based population genomics},
  journal = {Molecular Ecology},
  volume  = {28},
  number  = {21},
  pages   = {4737--4754},
  year    = {2019},
  doi     = {10.1111/mec.15253}
}

@inproceedings{yao2023react,
  title={{ReAct}: Synergizing Reasoning and Acting in Language Models},
  author={Yao, Shunyu and Zhao, Jeffrey and Yu, Dian and Du, Nan and Shafran, Izhak and Narasimhan, Karthik and Cao, Yuan},
  booktitle={International Conference on Learning Representations (ICLR)},
  year={2023}
}

@inproceedings{wang2024codeact,
  title={Executable Code Actions Elicit Better {LLM} Agents},
  author={Wang, Xingyao and Chen, Yangyi and Yuan, Lifan and Zhang, Yizhe and Li, Yunzhu and Peng, Hao and Ji, Heng},
  booktitle={Proceedings of the 41st International Conference on Machine Learning (ICML)},
  series={PMLR},
  volume={235},
  pages={50208--50232},
  year={2024}
}

@inproceedings{jimenez2024swebench,
  title={{SWE}-bench: Can Language Models Resolve Real-world Github Issues?},
  author={Jimenez, Carlos E. and Yang, John and Wettig, Alexander and Yao, Shunyu and Pei, Kexin and Press, Ofir and Narasimhan, Karthik R.},
  booktitle={The Twelfth International Conference on Learning Representations (ICLR)},
  year={2024}
}

@inproceedings{yang2024sweagent,
  title={{SWE}-agent: Agent-Computer Interfaces Enable Automated Software Engineering},
  author={Yang, John and Jimenez, Carlos E. and Wettig, Alexander and Lieret, Kilian and Yao, Shunyu and Narasimhan, Karthik R. and Press, Ofir},
  booktitle={Advances in Neural Information Processing Systems (NeurIPS)},
  volume={37},
  year={2024}
}

@inproceedings{wang2025openhands,
  title={{OpenHands}: An Open Platform for {AI} Software Developers as Generalist Agents},
  author={Wang, Xingyao and Li, Boxuan and Song, Yufan and Xu, Frank F. and Tang, Xiangru and Zhuge, Mingchen and Pan, Jiayi and Song, Yueqi and Li, Bowen and Singh, Jaskirat and Tran, Hoang H. and Li, Fuqiang and Ma, Ren and Zheng, Mingzhang and Qian, Bill and Shao, Yanjun and Muennighoff, Niklas and Zhang, Yizhe and Hui, Binyuan and Lin, Junyang and Brennan, Robert and Peng, Hao and Ji, Heng and Neubig, Graham},
  booktitle={The Thirteenth International Conference on Learning Representations (ICLR)},
  year={2025}
}

@inproceedings{zhang2024autocoderover,
  title={{AutoCodeRover}: Autonomous Program Improvement},
  author={Zhang, Yuntong and Ruan, Haifeng and Fan, Zhiyu and Roychoudhury, Abhik},
  booktitle={Proceedings of the 33rd ACM SIGSOFT International Symposium on Software Testing and Analysis (ISSTA)},
  pages={1592--1604},
  year={2024},
  doi={10.1145/3650212.3680384}
}

@inproceedings{xia2025agentless,
  title={Demystifying {LLM}-based Software Engineering Agents},
  author={Xia, Chunqiu Steven and Deng, Yinlin and Dunn, Soren and Zhang, Lingming},
  booktitle={Proceedings of the 33rd ACM SIGSOFT International Symposium on the Foundations of Software Engineering (FSE)},
  year={2025},
  doi={10.1145/3715754}
}

@article{zhou2024autoba,
  title={An {AI} Agent for Fully Automated Multi-Omic Analyses},
  author={Zhou, Juexiao and Zhang, Bin and Li, Guowei and Chen, Xiuying and Li, Haoyang and Xu, Xiaopeng and Chen, Siyuan and He, Wenjia and Xu, Chencheng and Liu, Liwei and Gao, Xin},
  journal={Advanced Science},
  volume={11},
  number={44},
  pages={e2407094},
  year={2024},
  doi={10.1002/advs.202407094}
}

@article{huang2025biomni,
  title={{Biomni}: A General-Purpose Biomedical {AI} Agent},
  author={Huang, Kexin and Zhang, Serena and Wang, Hanchen and Qu, Yuanhao and Lu, Yingzhou and Roohani, Yusuf and Li, Ryan and Qiu, Lin and Li, Gavin and Zhang, Junze and Yin, Di and Marwaha, Shruti and Carter, Jennefer N. and Zhou, Xin and Wheeler, Matthew and Bernstein, Jonathan A. and Wang, Mengdi and He, Peng and Zhou, Jingtian and Snyder, Michael and Cong, Le and Regev, Aviv and Leskovec, Jure},
  journal={bioRxiv},
  year={2025},
  doi={10.1101/2025.05.30.656746}
}

@inproceedings{schmidgall2025agentlab,
  title     = {Agent Laboratory: Using {LLM} Agents as Research Assistants},
  author    = {Schmidgall, Samuel and Su, Yusheng and Wang, Ze and Sun, Ximeng and Wu, Jialian and Yu, Xiaodong and Liu, Jiang and Moor, Michael and Liu, Zicheng and Barsoum, Emad},
  booktitle = {Findings of the Association for Computational Linguistics: EMNLP 2025},
  pages     = {5977--6043},
  year      = {2025},
  publisher = {Association for Computational Linguistics},
  doi       = {10.18653/v1/2025.findings-emnlp.320}
}

@article{lu2024aiscientist,
  title   = {The {AI} Scientist: Towards Fully Automated Open-Ended Scientific Discovery},
  author  = {Lu, Chris and Lu, Cong and Lange, Robert Tjarko and Foerster, Jakob and Clune, Jeff and Ha, David},
  journal = {arXiv preprint arXiv:2408.06292},
  year    = {2024}
}

@article{bran2024chemcrow,
  title   = {Augmenting large language models with chemistry tools},
  author  = {Bran, Andres M. and Cox, Sam and Schilter, Oliver and Baldassari, Carlo and White, Andrew D. and Schwaller, Philippe},
  journal = {Nature Machine Intelligence},
  volume  = {6},
  number  = {5},
  pages   = {525--535},
  year    = {2024},
  doi     = {10.1038/s42256-024-00832-8}
}

@article{ding2025scitoolagent,
  title   = {{SciToolAgent}: a knowledge-graph-driven scientific agent for multitool integration},
  author  = {Ding, Keyan and Yu, Jing and Huang, Junjie and Yang, Yuchen and Zhang, Qiang and others},
  journal = {Nature Computational Science},
  volume  = {5},
  pages   = {962--972},
  year    = {2025},
  doi     = {10.1038/s43588-025-00849-y}
}

@article{mehandru2025bioagents,
  title   = {{BioAgents}: Bridging the gap in bioinformatics analysis with multi-agent systems},
  author  = {Mehandru, Nikita and Hall, Amanda K. and Melnichenko, Olesya and Dubinina, Yulia and Tsirulnikov, Daniel and Bamman, David and Alaa, Ahmed and Saponas, Scott and Malladi, Venkat S.},
  journal = {Scientific Reports},
  volume  = {15},
  pages   = {39036},
  year    = {2025},
  doi     = {10.1038/s41598-025-25919-z}
}

\end{document}